\documentclass[prl,twocolumn,superscriptaddress]{revtex4-2}
\usepackage{amsmath,amssymb,mathrsfs}
\usepackage{graphicx}
\usepackage{colortbl}
\usepackage{CJK}
\begin{document}
\begin{CJK*}{UTF8}{bsmi}
\title{Gapless Spin Liquid and Non-local Corner Excitation in the Spin-$1/2$ Heisenberg Antiferromagnet on Fractal}
\author{Haiyuan Zou  (\CJKfamily{gbsn}邹海源)}
\altaffiliation{hyzou@phy.ecnu.edu.cn}
\affiliation{Key Laboratory of Polar Materials and Devices (MOE), School of Physics and Electronic Science, East China Normal University, Shanghai 200241, China}

\author{Wei Wang (\CJKfamily{gbsn}王巍)}
\affiliation{Tsung-Dao Lee Institute, Shanghai Jiao Tong University, Shanghai 200240, China}

\begin{abstract}
Motivated by the mathematical beauty and the recent experimental realizations of fractal systems, we study the spin-$1/2$ antiferromagnetic Heisenberg model on a Sierpi\'nski gasket. The fractal porous feature generates new kinds of frustration to exhibit exotic quantum states. Using advanced tensor network techniques, we identify a quantum gapless-spin-liquid ground state in fractional spatial dimension. This fractal spin system also demonstrates nontrivial non-local properties. While the extremely short-range correlation causes a highly degenerate spin form factor, the entanglement in this fractal system suggests a long-range scaling behavior. We also study the dynamic structure factor and clearly identify the gapless excitation with a stable corner excitation emerged from the ground-state entanglement.
Our results unambiguously point out multiple essential properties of this fractal spin system, and open a new route to explore spin liquid and frustrated magnetism.
\vspace{0.2cm}
\end{abstract}

\maketitle
\end{CJK*}
Quantum spin liquid is an exotic paramagnetic state defeating long-range orders but with nontrivial long-range entanglements and fractional excitations even at the zero temperature limit~\cite{ANDERSON1987,Savary2016,RMP2017SL}. It has been viewed as an insulating ground for generating doped high temperature superconductivity~\cite{RMP06Doping}, a fertile platform for studying quantum phases described by topological order~\cite{Wen2002order}, and a potential state for realizing quantum computing~\cite{Kitaev20032,RMP08Nayak}. 
Driven by all these interests, searching for spin liquids has attracted intensive and extensive attention. 
Theoretically, frustrated systems with strong quantum fluctuations are mostly targeted. They harbor competing behaviors at the boundary between different long-range ordered phases~\cite{Jiang2012J1J2,J1J2Plaq2014,Wang2016J1J2} and strange properties from peculiar geometries~\cite{Yan2011,Gapless2013,Liao2017Gapless,Wang22} or interactions~\cite{Kitaev20062}.
Inspiriting as it be, the nature of quantum fluctuation also bring obstacles (e.g., critical slowing down) for unbiased microscopic studies. Despite the great success of some exactly solvable models to establish the spin liquids, e.g., the honeycomb Kitaev model~\cite{Kitaev20062}, the nature of the ground states for many frustrated models is still under debate~\cite{Yan2011,Gapless2013,Liao2017Gapless,Jiang2012J1J2,J1J2Plaq2014,Wang2016J1J2}. Furthermore, the identification of exotic excitations of a spin liquid through its dynamical properties is an extremely difficult task~\cite{PRL2014DSF,PNAS2019Zhu}. These challenges push forward the exploration on new systems with unambiguous ground states of non-local properties and clear excitations to realize spin liquid physics.

While such exploration has been widely tested on integer dimensional lattice systems, recent experimental advances~\cite{NatChem2015Wu,NatPhy2019Smith,PRL2021Jia} on fractal lattice systems trigger various theoretical studies, e.g., on topological character~\cite{PRB2018T1,PRB2019T2,PRR2020Wang,Light2020Segev}, localization~\cite{PRB2016Sticlet,PRB2017Kosior}, and transport properties of electronic states~\cite{PRB2016Yuan,PRB2017Yuan,PRB2020Yuan}, etc. 
However, quantum magnetism on fractal systems, featuring an interplay between self-similarity and quantum fluctuation, still remains nascent and indicates exotic quantum phase of matter. 

\begin{figure}[th]
\hspace{-0.3cm}
\includegraphics[width=0.5\textwidth]{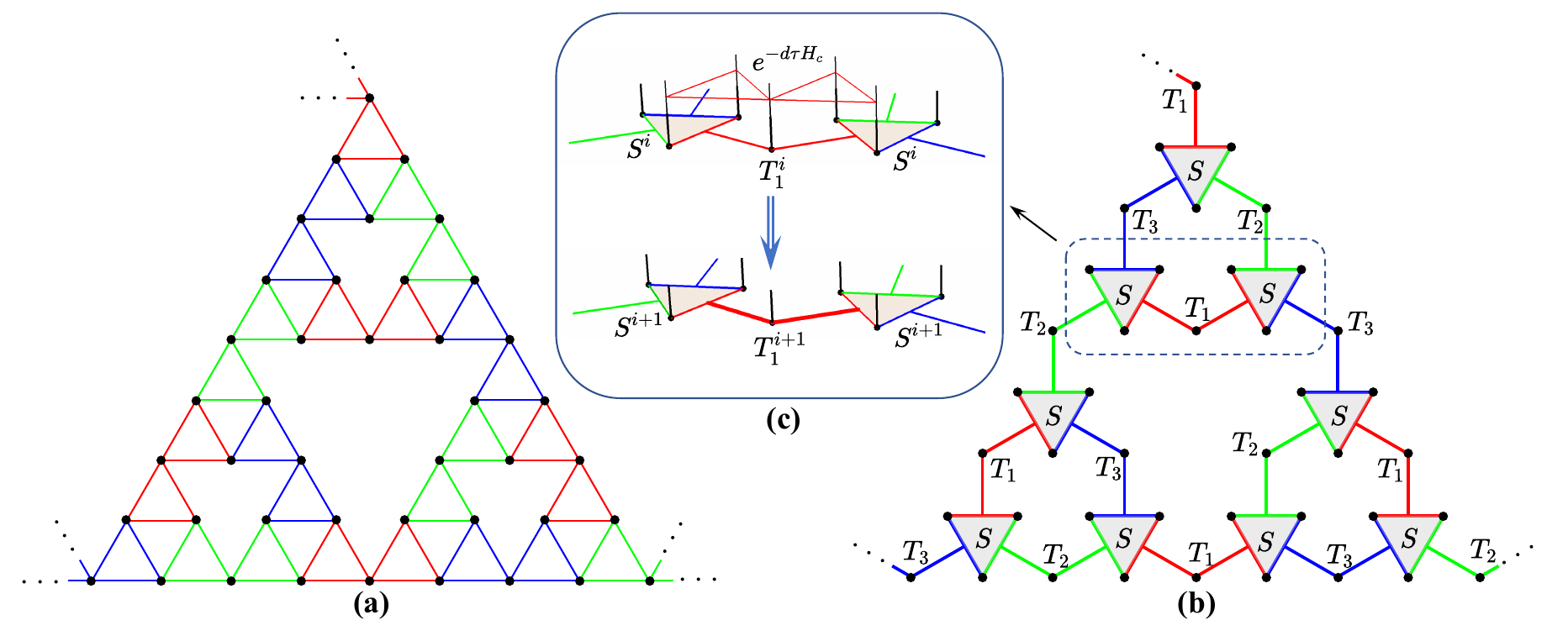}
\caption{Geometry and the tensor network representation of a Sierpi\'nski gasket. (a) An infinite Sierpi\'nski gasket with a highlight of its level-3 part is shown, where the dashed line represents the environment. The Hamiltonian of the total system is decomposed into three parts with each part contains two triangles, labeled by different colors. (b) The ground state wave function is represented by a tensor network repeating three local tensors $T_i$ and a simplex $S$ tensor. In each local tensor, a black dot represents a physical leg. (c) One step of the time evolution with the operator $e^{-d\tau H_c}$ for the dashing-bounded region in (b) is shown. Repeating this procedure will evolve the state to a stable ground state.}
\label{fig:model}
\end{figure}

A paradigmatic example of fractal lattice is the Sierpi\'nski gasket which can be generated iteratively as follows. The level-0 Sierpi\'nski gasket is a simple triangle. Then, from the level-$n$ structure [Fig~\ref{fig:model}(a) shows a snapshot of the level-3 Sierpi\'nski gasket], linking the three centers of the upper triangular bonds in each triangle generates the level-$n$+1 Sierpi\'nski gasket. This bond bifurcating operation triples the number of bonds and can be represented by a Hausdorff dimension $D_H$=$\ln(3)/\ln(2)\approx$1.585. Alternatively, the Sierpi\'nski gasket can be generated through removing lattice sites from an infinite triangular lattice. Actually, similar process can generate the well-studied Kagome lattice with a translationally symmetric porous pattern. However, the pore-doping process in the Sierpi\'nski gasket gives rise to a hierarchical porous pattern which underlies novel geometric frustration that breaks the translational symmetry, dubbed as ``porous frustration''. Such porous frustration suggests new paradigm of spin liquid.

The self-similar feature of the Sierpi\'nski gasket fractal lattice implies the feasibility of renormalization group (RG) methods. Indeed, the classical statistical models on fractal system are well studied using RG techniques~\cite{PRL1980Critical,Nishino2019}. Meanwhile, the past few decades witness the fast development of tensor network methods~\cite{PEPS1,iPEPS,TEBD,Schollwck2011,Ors2014,MERA,TRG,TERG}, stemmed from Density Matrix Renormalization Group~\cite{DMRG} and widely used for quantum many-body systems. It is plausible to apply tensor network methods to quantum spin models on fractal system, especially for the reason that the coarse-graining type tensor networks~\cite{TRG,HOTRG} provide a nature way to contract the environment of local operators in fractal systems. However, the large quantum fluctuation generated from the intrinsic porous frustration in the fractal lattice prevent conventional trivial updating scheme to the ground-state of quantum spin model on a Sierpi\'nski gasket. Therefore, although some schemes are proposed~\cite{EPJ2016Su}, unambiguous ground state feature are still at large. These difficulties also imply unusual entanglement behaviors of fractal systems need to be uncovered.  

Here, using a state-of-the-art tensor network in the thermodynamic limit, we obtain strong evidences of a gapless spin liquid as the ground state of the spin-$1/2$ Heisenberg antiferromagnetic (AF) model in the Sierpi\'nski gasket fractal lattice system. First, using a new time evolution scheme, local variables such as the energy and magnetization can be calculated, which identifies a stable disordered ground state with gapless scaling behaviors. Second, two non-local properties provide the short-range correlated but highly entangled feature of the spin liquid ground state. A large degeneracy in the static spin structure further confirms the disordered signature. A one dimensional entanglement entropy scaling behavior implies strong entanglement growth. Third, the properties suggested from the static properties are further confirmed by the dynamic structure factor study. Not only the bulk gapless excitation but also a stable corner excitation are clearly identified. The corner excitation is consistent with a intuitive physical picture of flipping non-local but highly entangled spins at three corners through the whole gapless spin liquid bulk in between. 
Our results not only provide unambiguous ground state properties of this fascinating system but also illustrate a strong entanglement behavior and the associated corner excitation, which broadly advance the field of spin liquid and the frustrated magnetism.

\begin{figure}[t]
\includegraphics[width=0.5\textwidth]{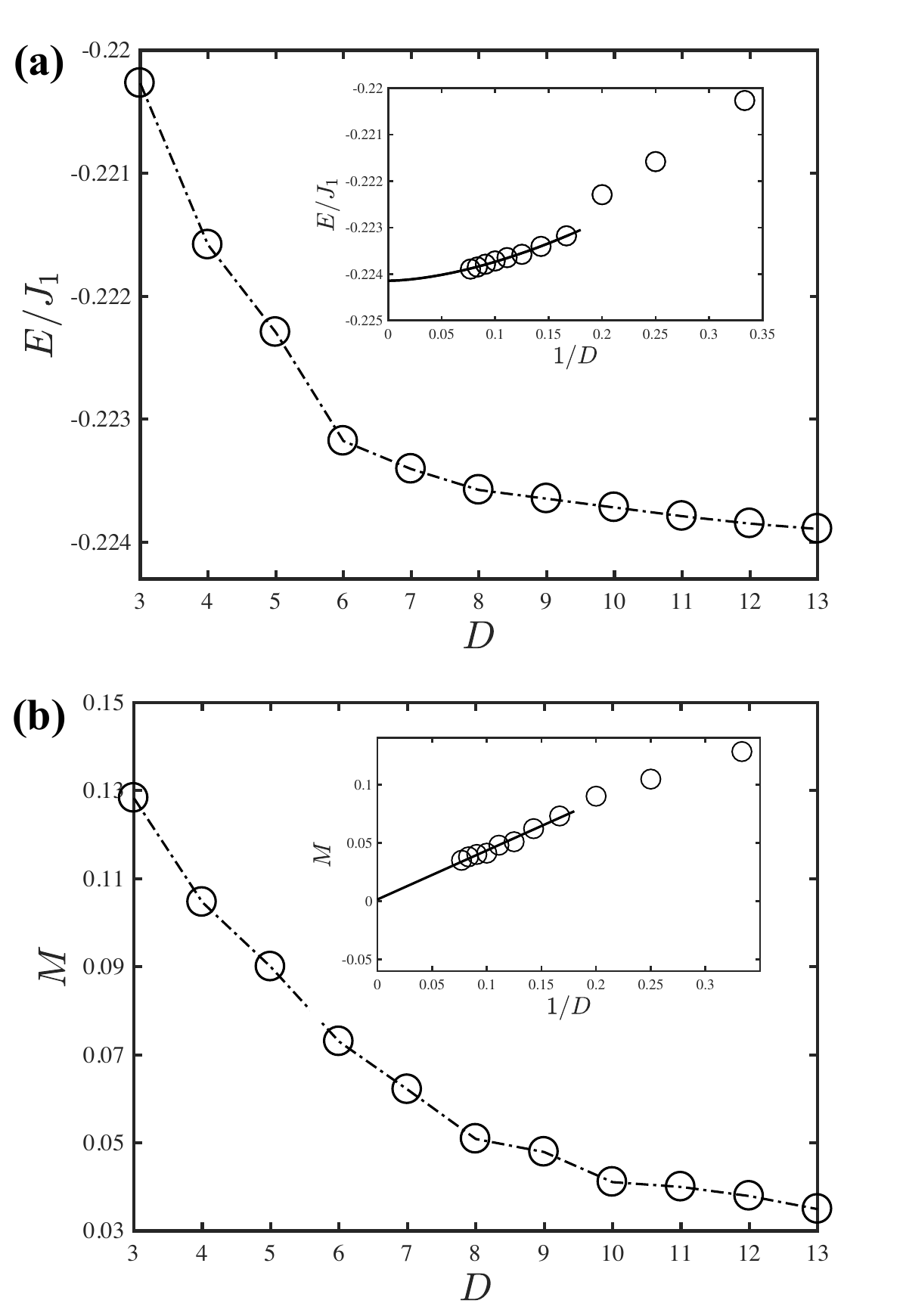}
\caption{Local properties of the system. (a) The ground state energies are converged as $D$ is increased. the insert shows the convergent behavior of the ground state energy per bond as function of $1/D$ with the form $c/D^{\alpha}+E_0$, where $c$=0.0193, $\alpha$=1.6743, and $E_0$=-0.2241(1). (b) The local magnetization per site as a function of $D$ is shown. The insert indicates that the $M$ is suppressed linearly as function of $1/D$ to zero when $D\rightarrow\infty$, with the form $c'/D+M_0$, where $c'$=0.4202 and $M_0$=0.0014(27).}
\label{fig:ME}
\end{figure}

In this work, we consider the simple spin-$1/2$ Heisenberg AF model on the Sierpi\'nski gasket lattice. it is described by the Hamiltonian:
\begin{equation}
H=J\sum_{\rm b}\mathbf{S_i}\cdot\mathbf{S_j},
\label{eq:model}
\end{equation}
where $\mathbf{S_i}$ is the spin operator on site $i$, $J>0$ is the nearest neighbors AF exchange couplings on a Sierpi\'nski lattice. In the tensor network calculation, the tensor network wave function is constructed by building blocks with three different local tensors $T$s and one $S$, shown in Fig.~\ref{fig:model}(b). Specifically, the $S$ with a simplex structure~\cite{PESS} represents the smallest down-triangle in the fractal. To deal with the frustration generated from the loop structure of the self-similar triangles, an efficient updating scheme is highly demanded. We find that regrouping the local Hamiltonian into a double triangular structure, shown in Fig.~\ref{fig:model}(a), can successfully avoid locally entangled update. Starting from a random configuration of local tensors, the tensor network converges to a stable ground state by the iterative imaginary time evolution updating scheme with the operator $e^{-d\tau H_c}$, which employs the double triangular local Hamiltonian $H_c$ [shown in Fig.~\ref{fig:model}(c)]. After convergence of the ground state, physical variables for the fractal system in the thermodynamic limit can be calculated by coarse-graining contraction of the local tensors which have no increases of the virtual bond dimension $D$. Therefore, without the approximation of contraction, $D$ is the only tuning parameter during the tensor network calculation for a Sierpi\'nski gasket~\cite{noteS}.  

The results of the ground-state energy per bond $E$ and the local magnatization $M$ of the Sierpi\'nski gasket Heisenberg antiferrimagnet in the themodynamical limit are shown in Fig.~\ref{fig:ME} as a function of $D$. In Fig.~\ref{fig:ME}(a), $E$ converges algebraically with $D$, indicating a gapless ground state~\cite{PRB2012Pirvu,Liao2017Gapless}. The power-law form $E(D)=c/D^{\alpha}+E_0$ shown in the insert gives the extrapolated ground-state energy at $D\rightarrow\infty$ with $E_0=-0.2241(1)$. As expected, in Fig.~\ref{fig:ME}(b), $M$ is suppressed algebraically as $D$ increases. A linear extrapolation as function of $1/D$ shown in the insert gives nearly vanished $M$ [$M(D\rightarrow\infty)=0.0014(27)$], suggesting a paramagnetic ground state. Therefore, the combining results of local quantities suggest a gapless paramagnetic ground state of the Sierpi\'nski gasket Heisenberg antiferrimagnet. 

To further investigate the ground state properties of the Sierpi\'nski gasket Heisenberg antiferrimagnet, we also preform calculation on non-local quantities. Due to the special porous fractal structure, the signatures of non-local quantities need to be illustrated. It is inefficient to define a boundary line of the bulk on a fractal structure. For example, a infinite Sierpi\'nski gasket has zero area but infinite perimeter $l$  [the sum of all the bonds in the Level-$n$ Sierpi\'nski gasket $l(n)\sim (3/2)^n$]. However, due to the self-similarity, segment lines contains fixed number of sites at anywhere of the Sierpi\'nski gasket are identical with each other except at the corners. Additionally, the porous structure breaks the two-dimensional (2D) periodicity and there is no well-defined 2D Brillouin zone. Therefore, considering the correlation along a line and the entanglement between the line with its environment in the Sierpi\'nski gasket give proper non-local properties of the system.

\begin{figure}[h]
\includegraphics[width=0.48\textwidth]{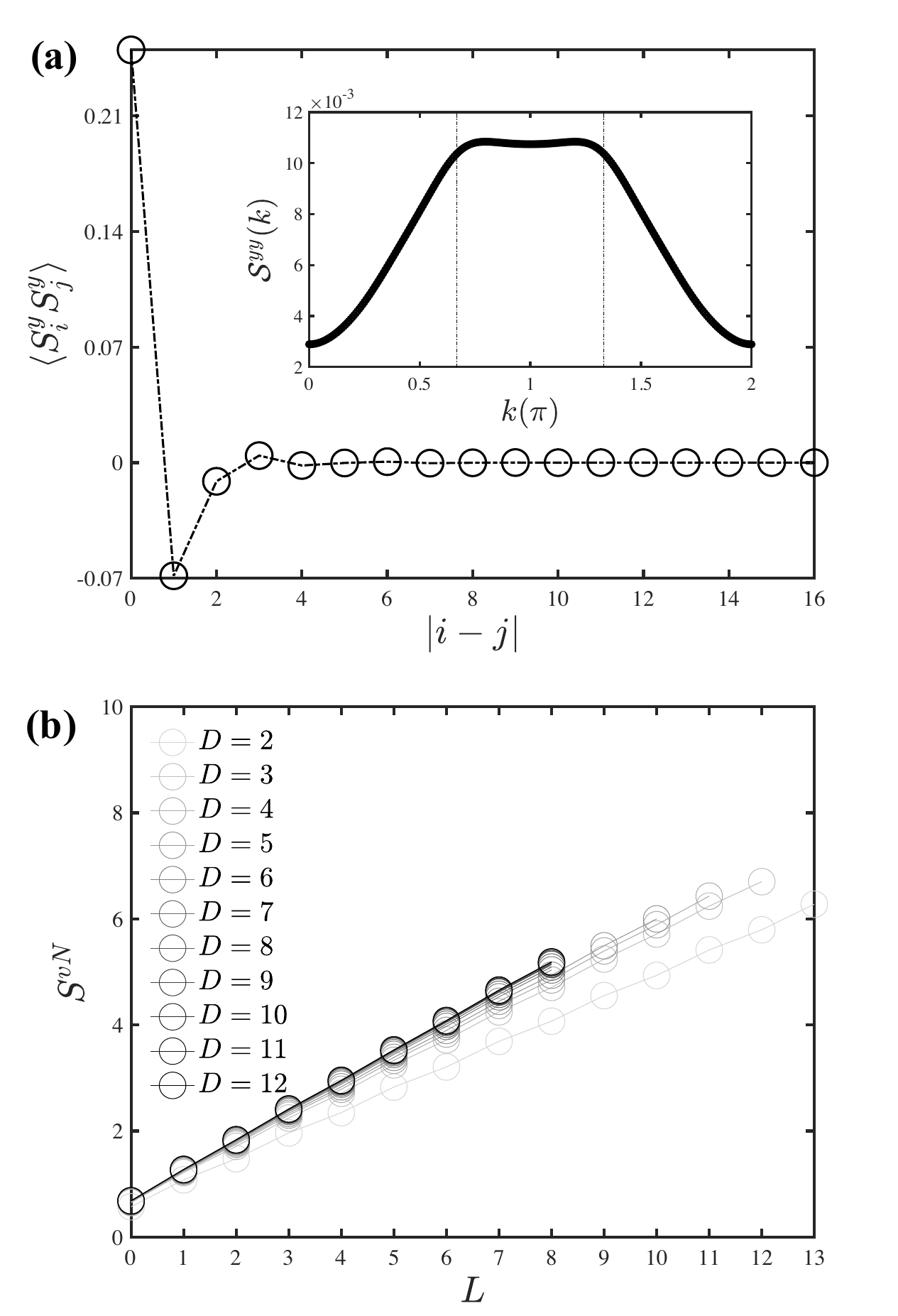}
\caption{Non-local properties of the system. (a) The spin-spin correlation along a line shows a fast decay, suggests a short-range correlation. The insert shows that the spin form factor $\mathcal{S}^{yy}(k)$ with a flat plateau in between two vertical dashed lines at $k=2/3\pi$ and $4/3\pi$, the momentum corresponding to a 120$^\circ$ order. (b) The entanglement entropy as a function of distance at different $D$ shows a one-dimensional volume law growing behavior with the form $aL+\gamma$. At $D$=12, $a$=0.56, and $\gamma$=0.70$\approx\log(2)$.} 
\label{fig:static}
\end{figure}

The first non-local quantity, the spin-spin correlation $\langle S^\mu_iS^\mu_j\rangle$ ($\mu=x,y,z$) as a function of the distance between two sites $i$ and $j$, is calculated. Figure~\ref{fig:static}(a) shows a fast decay behavior of $\langle S^y_iS^y_j\rangle$ as the distance is increased, which indicates very short-range correlation in this fractal system. The static spin structure factor $\mathcal{S}^{yy}(k)$ shown in the insert, obtained from a Fourier transform from the correlation to the $k$-space, has a highly degenerate plateau at $2\pi/3\lesssim k\lesssim 4\pi/3$ without any clear peak and provides evidence of the disorder feature of the ground state. 

Secondly, we also calculate the von Neumann entanglement entropy $S^{vN}(L)=-\rm{Tr}\rho_{A(L)}\ln\rho_{A(L)}$ for a subsystem $A$ with a line of $L$ spins, where $\rho_A\equiv\rm{Tr}_B|\Psi_A\otimes\Psi_B\rangle\langle\Psi_A\otimes\Psi_B|$ is the reduced density matrix of subsystem $A$ with a traced environment $B$. 
The results of $S^{vN}(L)$ at different $D$ is shown in Fig.~\ref{fig:static}(b). As $D$ increases, $S^{vN}(L)$ saturates to a line ($aL+\gamma$), following one-dimensional (1D) volume law with $\gamma\sim\log2$, different from the Kitaev-Preskill scheme with a topological order~\cite{TEE2006}.  

These non-local properties, especially the entangled behavior in the fractal spin system, imply different kind of excitations from the conventional valence bond picture. In a Sierpi\'nski gasket, strong porous frustration can suppress the formulation of valence bonds. Instead of breaking the valence bond to form quasiparticles, gapless excitations will take place. Although there is no clear bulk-boundary correspondence in the fractal system, corner states of the fractal system can entangled with each other to generate nontrivial excitations.

To justify the gapless excitation suggested by our calculation and detect possible corner excitations, 
we study the dynamic properties of the fractal system. 
With the implementation of the tensor network ground state wavefunction, the space-time correlation $\langle S^\mu(j,t)S^\nu(i,0)\rangle$ is calculated, where $\mu,\nu=x,y,z$ and $i,j$ are two sites along a line. The dynamic structure factor (DSF)
\begin{equation}
\mathcal{S}^{\mu\nu}(\omega, k)=\int dt\sum_{j-i}e^{i[\omega t-k\cdot(j-i)]}\langle S^\mu(j,t)S^\nu(i,0)\rangle,
\end{equation}
is obtained correspondingly. The DSF is related with the intensity measurement in experiments, e.g., Inelastic neutron scattering. For a 1D magnetic system, the DSF results from tensor network methods can precisely capture the nontrivial excitation information and give clear guide for the spectrum measurement in experiments on magnetic materials~\cite{Zou2021PRL}. To investigate both the bulk excitation and possible corner mode simultaneously, we consider a large finite fractal lattice system (the level-4 Sierpi\'nski gasket with 123 sites) in the DSF calculation~\cite{noteS}. 

\begin{figure}[h]
\includegraphics[width=0.5\textwidth]{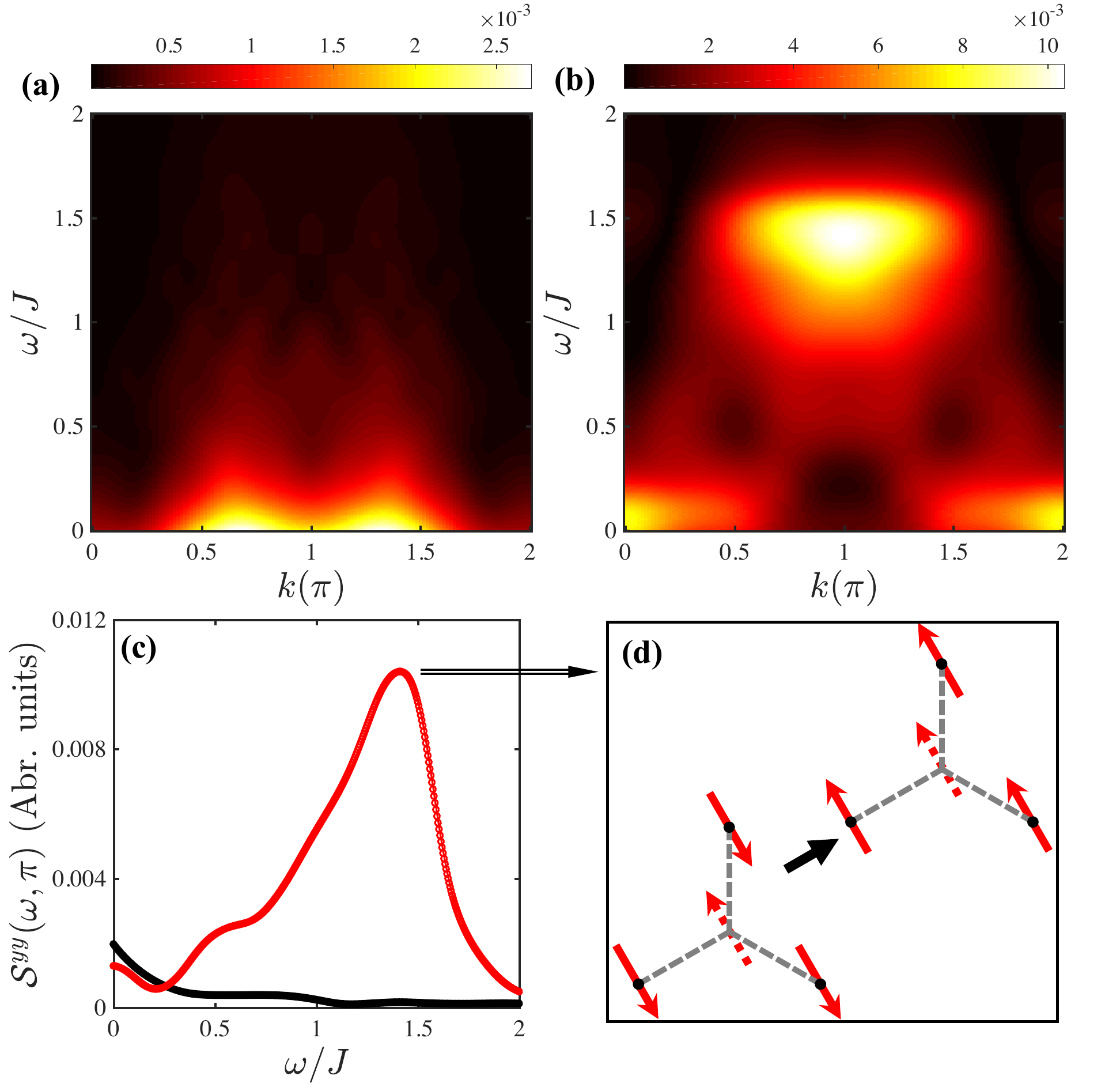}
\caption{Dynamic properties of the system. (a) The DSF of the system, shows gapless continuous excitations at in a large momentum region. (b) The corner excitation at $\omega\sim 1.5J$ can be found in the middle $k$ region. (c) The comparison of the bulk gapless excitation and corner excitation at $k=\pi$. (d) The corner excitation at (c) is demonstrated as a consequence from a spin flipping process of the non-local three corners, which suggests highly entanglement among the three corners.}
\label{fig:dsf}
\end{figure}

The DSF intensity result with the operator $S^{\nu}(i,0)$ inserted at site $i$ inside the fractal lattice provides the general information of the large fractal system. In Fig.~\ref{fig:dsf}(a), the DSF $\mathcal{S}^{yy}(\omega, k)$ in the $\omega$-$k$ plane gives continuous strong intensity at $\omega =0$ in a large region of $k$, which provides a smoking gun evidence of the gapless excitation suggested by the static quantity calculations. To detect the possible corner excitation, we place the operator $S^{\nu}(i,0)$ at one of the three corners of the finite fractal system. Surprisingly, the resulting DSF gives a significant peak at $\omega\sim 1.5J$, indicating a clear gapped excitation of the corner mode, shown in Fig.~\ref{fig:dsf}(b). The comparison of DSF intensity in between the bulk gapless excitation and the gapped corner mode at $k=\pi$, shown in Fig.~\ref{fig:dsf}(c), suggests that the corner excitation is an additive mode.
 
This unusual corner excitation can be elucidated by a simple physical picture [Fig.~\ref{fig:dsf}(d)] resulted from the ground state entanglement of the fractal system. Spins at the three corners are non-local degrees of freedom connected by the ground state entanglement through the whole gapless spin liquid fractal system. A local spin flip at one corner strongly entangles the whole fractal system, can generate spin flips at spins on the other two corners simultaneously. Considering the middle bulk gapless spin liquid [dashed lines in panel (d)] as an effective spin to lower the energy of the whole system with the three corners, this spin flip process gives rise to a gap $\omega=1.5J$, coincident with the corner excitation obtained in panel (b) and (c).
    
The significant excitation properties and the associated highly entangled nature of the Sierpi\'nski gasket spin system make the fractal system a new platform to explore spin liquid physics. Unique feature of fractal systems different from conventional integer dimensional systems need to be further investigated. Our numerical evidence of the entanglement entropy scaling can motivate theoretical study of possible topological entanglement~\cite{TEE2006} in fractal by choosing different subsystem instead of a line segment. The DSF results can guide future experiments on the manipulation of fractal systems. The gapless-spin-liquid ground state found in fractal systems might be considered as a potential state to emerge a chiral spin liquid~\cite{PNAS2019Zhu}. Future non-equilibrium study of the fractal systems may open new route to realizing topological quantum computing~\cite{RMP08Nayak}.

In summary, using a state-of-the-art tensor network ansatz, we identify a gapless spin liquid ground state in the spin-$1/2$ AF Heisenberg model on the Sierpi\'nski fractal system for the first time. Results on local quantities and correlations indicate the gapless and disorder features. An example of the entangled behaviors in this fractal system are demonstrated. Furthermore, the dynamic studies on excitations give unambiguous signals of the bulk gapless excitation and a stable non-local corner excitation followed by a intuitive physical picture suggested from the ground state entanglement. Our results based on advanced techniques on a fractal system open a new direction to probe spin liquid physics and frustrated systems in general.

We thank Ruizhen Huang, Haijun Liao, and Tao Xiang for helpful discussions. This work is supported by National Natural Science Foundation of China Grant No.~12274126 (H.Z.). Part of this work was done during the virtual program ``Tensor Networks in Many Body and Quantum Field Theory" held at the Institute for Nuclear Theory, University of Washington, Seattle (INT 21-1c). 


\pagebreak

\newpage
\widetext
\begin{center}
\textbf{\large Supplemental Materials for ``Gapless Spin Liquid and Non-local Corner Excitation in the Spin-$1/2$ Heisenberg Antiferromagnet on Fractal"}
\end{center}

\setcounter{equation}{0}
\setcounter{figure}{0}
\setcounter{table}{0}
\makeatletter
\renewcommand{\thefigure}{S\arabic{figure}}
\renewcommand{\thetable}{S\arabic{table}}
\renewcommand{\theequation}{S\arabic{equation}}
\renewcommand{\bibnumfmt}[1]{[S#1]}
\renewcommand{\citenumfont}[1]{S#1}
\makeatother

\section{Self-Similarity in Tensor Networks}

The local tensor wave-function of the Sierpi\'nski Gasket fractal spin system is constructed by a tensor network ansatz with a simplex structure~\cite{PESSS}. In the main text, to avoid state unstablity due to the entanglement from the loops of triangles in Sierpi\'nski structure, a time-evolution procedure with a double-triangle structure is applied, which successfully converges the tensor network state to the ground state. In this study, simple update considering the bond vector $\lambda$s, which connect $S$ and $T$ tensors, is used. On the coarse-graining step of calculating physical quantities, $\lambda$s can be easily contracted into either $S$ or $T$ tensors, therefore, we omit $\lambda$s in the following illustration. 

\begin{figure}[h]
\centering
\includegraphics[width=0.8\textwidth]{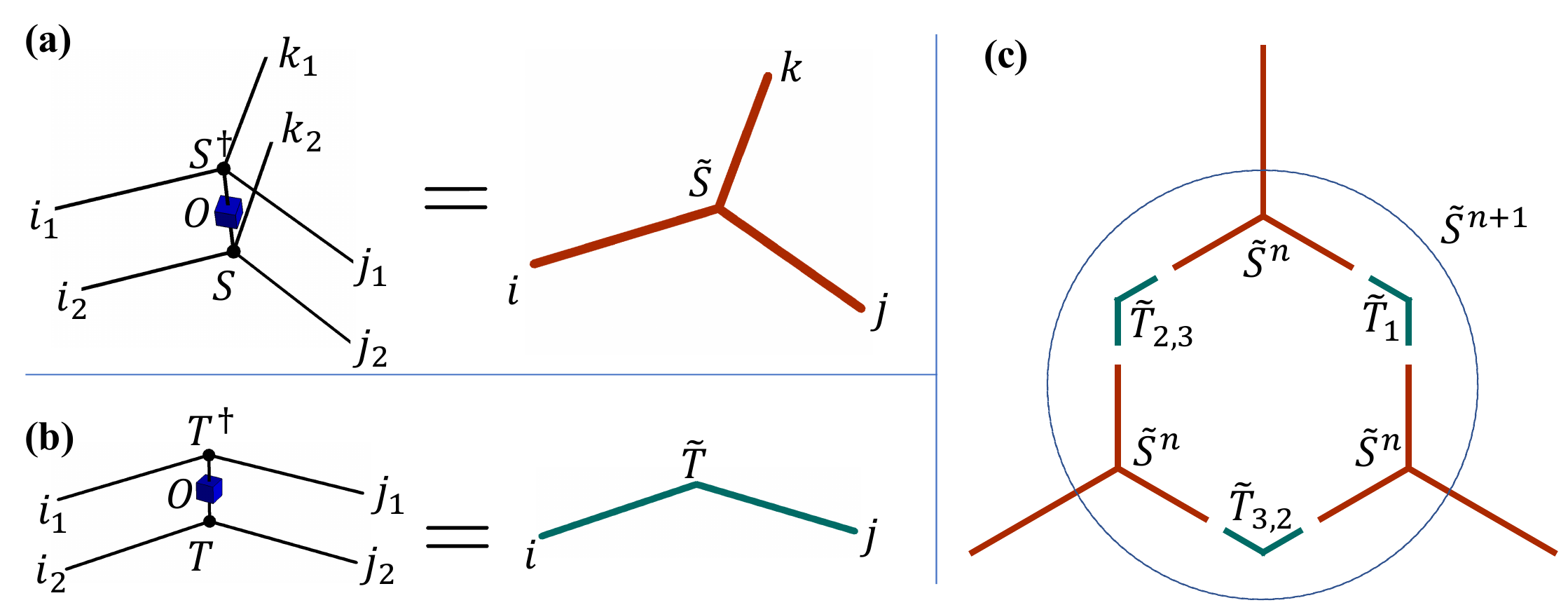}
\caption{Coarse-graining steps. (a) $\tilde{S}$ is constructed from contracting the real indices of $S$ with the operator $O$ inserted, similar step is shown for $\tilde{T}$ in (b). (c) The tensor contraction step is shown in the circle. The new $\tilde{S}^{n+1}$, resulted from $\tilde{S}^{n}$ and $T$s, not only has the same structure but only the same virtual bond dimension with $\tilde{S}^{n}$. The positions of $T_2$ and $T_3$ are determined by the odevity of the integer $n$. E.g., for odd $n$, $T_1$, $T_2$, and $T_3$ are placed anticlockwise in the circle.}
\label{fig:structure}
\end{figure}

To calculate local or non-local physical quantities $O$, the physical indices is contracted first with the local $T$ or $S$ tensor (for a simple norm calculation of the local tensor, $O$ is just a identity matrix), shown as 
 \begin{eqnarray}\nonumber
\tilde{S}_{i,j,k}&=&S^\dagger_{i_1,j_1,k_1,l_1}O_{l_1,l_2}S_{i_2,j_2,k_2,l_2}\\
\tilde{T}_{i,j}&=&T^\dagger_{i_1,j_1,l_1}O_{l_1,l_2}T_{i_2,j_2}
\label{eq:contract}
 \end{eqnarray}
where $i=\{i_1,i_2\}$, $j=\{j_1,j_2\}$, and $k=\{k_1,k_2\}$ are the regrouped tensor links with bond dimension $D^2$, in which $D$ is the virtual bond dimension of the original $T$ or $S$ tensor.
The contraction steps are shown in Fig.~\ref{fig:structure}(a,b). the resulting $\tilde{S}$ and $\tilde{T}$ are used in the following steps.

The self-similar structure of the fractal system can be captured by the tensor network structure. In the coarse-graining step shown in Fig.~\ref{fig:structure} (c), after one iteration step of the tensor contraction from $\tilde{S}^n$ and $\tilde{T}$s, the resulting $\tilde{S}^{n+1}$ has the exact same structure with $\tilde{S}^n$. Remarkably, different from a 2D tensor contraction procedure, there is no increase of virtual bond dimension of $\tilde{S}^{n+1}$, which makes the coarse-graining step to the thermodynamic limit without further truncation approximation for a fixed $D$.

\section{calculation of Static physical quantities}

\begin{figure}[h]
\includegraphics[width=0.9\textwidth]{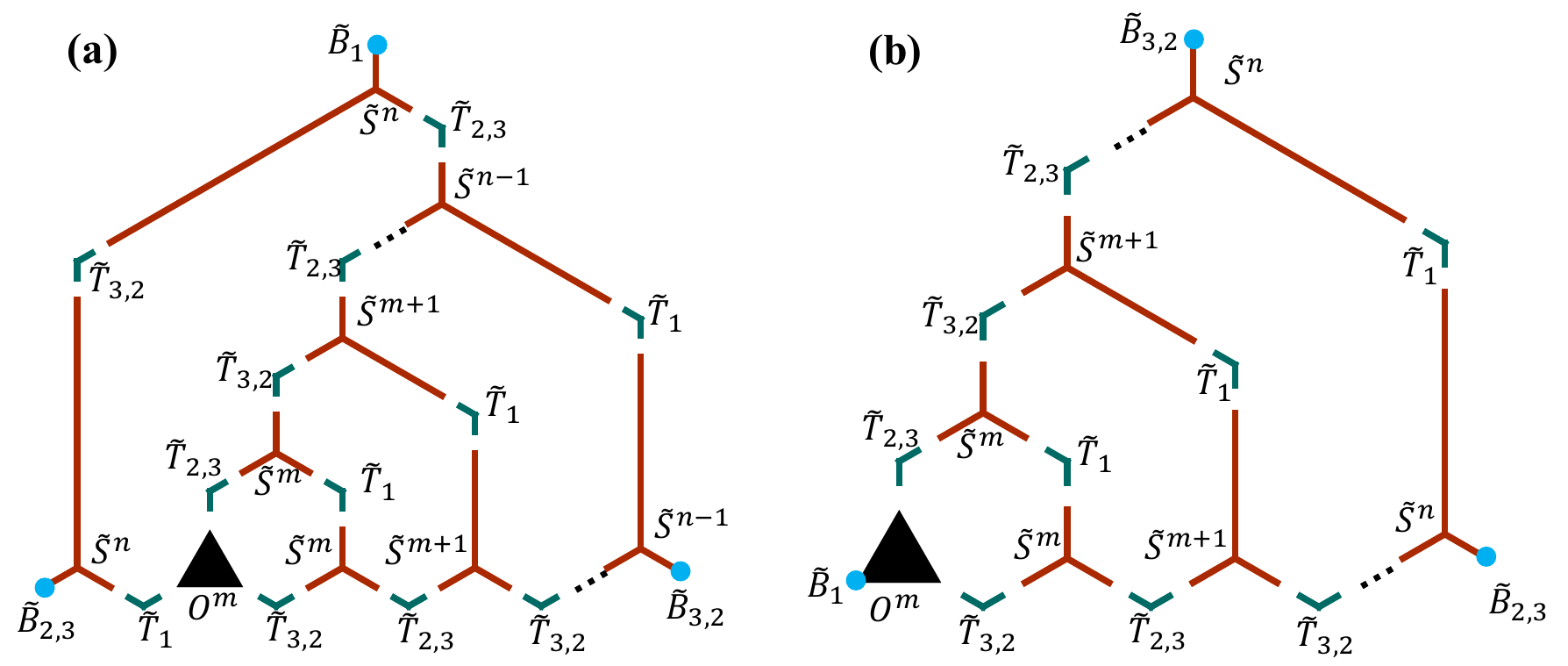}
\caption{The procedure of calculating the expectation values of local operators in the thermodynamic limit. (a) expectation values at the center of the fractal system. the black triangle represents any physical operator to be calculated. The tensor network environment is constructed by $\tilde{T}$s, calculated $\tilde{S}^n$ and corner tensor $\tilde{B}$s. (b) the same procedure for an operator placed on the corner. The $\tilde{T}$ surrounding $\tilde{S}^n$ is placed clockwise for odd $n$ and anticlockwise for even $n$.}
\label{fig:contraction}
\end{figure}

In the main text, all the static variables, e.g. ground state energy per bond $E$, local magnetization $M$, the static spin correlation functions $S^y_iS^y_j$, and the entanglement entropy, are calculated in the thermodynamic limit. Figure~\ref{fig:contraction}(a,b) show two procedures of expectation value calculations. Due to the self-similar structure of the tensor networks, the results of static variables are not sensitive to the locations of the positions of the operators and the corner effects can be neglected. In the main text, the results of static variables are calcualted through the procedure shown in Fig.~\ref{fig:contraction}(a). In our calculation, we put the vector $\tilde{B}$, constructed by a self contraction of $T$ tensors, at the three corners. The self-contraction is shows as $\tilde{B_a}_{,j}={T_a}^\dagger_{\{i_1,j_1,p_1\}}{T_a}_{\{i_1,j_2,p_1\}}$, where the physical indices $p_1$ and the outside virtual indices $i_1$ are contracted, and $j=\{j_1,j_2\}$ for $a=1,2,3$.  

Specifically, the ground state energy and local magnetization shown in Fig.~2 in the main text are the averaging results from a level-2 (with 15 sites and 27 bonds) Sierpi\'nski gasket placed into the tensor environment shown in Fig.~\ref{fig:contraction}(a). Similarly, the spin correlation and entanglement entropy results shown in Fig.~3 in the main text are calculated by putting a level-4 Sierpin\'nski gasket into the black triangle in Fig.~\ref{fig:contraction}(a).

\begin{figure}[h]
\includegraphics[width=0.7\textwidth]{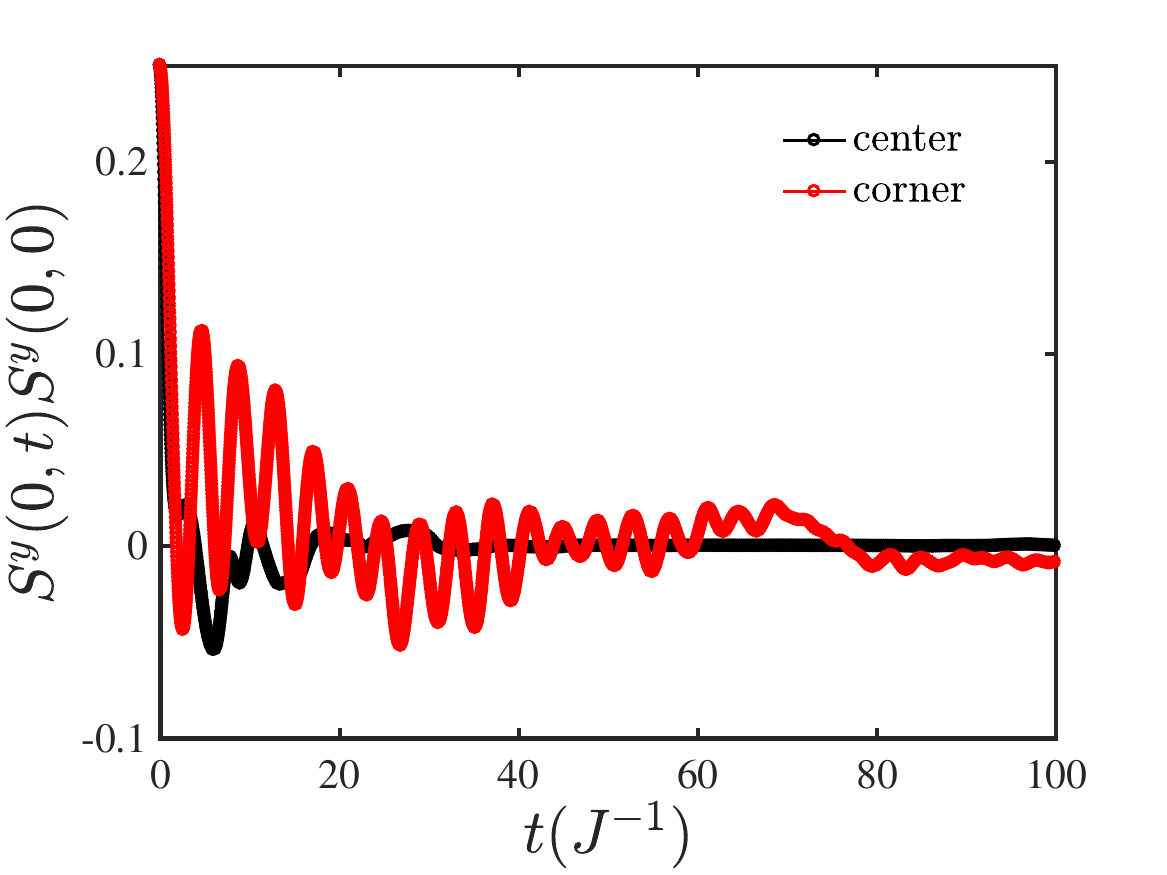}
\caption{The zero distance spin correlation function as a function of evolution time for putting the spin at different locations. The black dots show a damping correlation when the spin is placed at the center of a Sierpi\'nski gasket. The red dots show a high-frequency oscillation of the correlation when it is placed at one of the corners.}
\label{fig:compare}
\end{figure}

\section{Space-time correlation}

Different from the static variables, the time-dependent spin-spin correlation results are dependent on the localtion of spins. If one of the spin is put at a corner, the time evolution of the spin-spin correlations is modified by the corner effect. Figure~\ref{fig:compare} shows the significant difference of the zero distance spin correlation results between the bulk and the corner. If a spin is placed at one of the three corners of the fractal system, the correlation have a high energy oscillating mode contributed from the corner effect. To further investigate this corner effect, The full space-time correlation on a finite Sierpi\'nski gasket is calculated. In Fig.~4 of the main text, results are calculated on a level-4 Sierpi\'nski gasket with 123 sites. 

Starting from the obtained ground state wave-function, the space-time correlation can be calculated by a real time evolution~\cite{White2008SpectrumS}. In the correlation calculation, we have the virtual bond dimension $D=6$, the discretized time interval for the evolution is fixed as $\tau=0.02J^{-1}$, and the interating step number is upto 5000. The dynamical structure factor is then obtained by the Fourier transform of the space-time correlation with a broad window~\cite{White2008SpectrumS}.

\end{document}